%% file: mmWave_ICASSP.tex
\documentclass{article}
\input{input.tex}

\newcommand{\sref}[1]{{Section}~\ref{#1}}
\usepackage{epstopdf}
\usepackage{enumerate}
\usepackage{algorithmicx}
\usepackage{algorithm}
\usepackage{amsmath}
\usepackage[noend]{algpseudocode}
\usepackage{float}
\usepackage{color}
\usepackage{makeidx}
\usepackage{bbm}
\usepackage{spconf,graphicx}
\usepackage{cite}
\allowdisplaybreaks

\def\j{\mathrm{j}}
\begin{document}
\title{Compressed Sensing Based Multi-User Millimeter Wave Systems: \\ How Many Measurements Are Needed?}
\name{Ahmed Alkhateeb$^{\dag}$, Geert Leus$^{\ddag}$, and Robert W. Heath Jr.$^{\dag}$ \thanks{This work is supported in part by the National Science Foundation under Grant No. 1218338 and 1319556, and by a gift from Huawei Technologies, Inc.}}
\address{$^\dag$ The University of Texas at Austin, TX, USA, Email: $\{$aalkhateeb,  rheath$\}$,@utexas.edu \\
$^\ddag$ Delft University of Technology, The Netherlands, Email: {g.j.t.leus@tudelft.nl} }

\maketitle

\begin{abstract}
Millimeter wave (mmWave) systems will likely employ directional beamforming with large antenna arrays at both the transmitters and receivers. Acquiring channel knowledge to design these beamformers, however, is challenging due to the large antenna arrays and small signal-to-noise ratio before beamforming. In this paper, we propose and evaluate a downlink system operation for multi-user mmWave systems based on compressed sensing channel estimation and conjugate analog beamforming. Adopting the achievable sum-rate as a performance metric, we show how many compressed sensing measurements are needed to approach the perfect channel knowledge performance. The results illustrate that the proposed algorithm requires an order of magnitude less training overhead compared with traditional lower-frequency solutions, while employing mmWave-suitable hardware. They also show that the number of measurements need to be optimized to handle the trade-off between the channel estimate quality and the training overhead.
%
\end{abstract}
\section{Introduction} \label{sec:intro}
Millimeter wave (mmWave) communication  is a promising technology for future cellular systems~\cite{pi2011,Rapp5G,Cov_Magazine,Boccardi}. Directional precoding with large antenna arrays appears to be inevitable to support longer outdoor links and to provide sufficient received signal power. The design of precoding matrices, though, is usually based on complete channel state information, which is difficult to achieve in mmWave due to (i) the large number of antennas, (ii) the small signal-to-noise ratio (SNR) before beamforming, and (iii) the different hardware constraints which impacts the training signal design \cite{mmWave_Estimation_2013}. Therefore, developing low-complexity mmWave channel estimation techniques is crucial for the mmWave system operation.

To overcome the hardware limitations, analog beamforming solutions were proposed in~\cite{Wang1, .13c, chen2011multi, Multilevel, Tsang}, which rely on controlling the phase of the signal transmitted by each antenna via a network of analog phase shifters. To avoid the need for explicit channel knowledge, the algorithms in ~\cite{Wang1, .13c, chen2011multi, Multilevel, Tsang} depend on beam training which iteratively designs the analog beamforming coefficients at the transmitter and receiver. These solutions, however, have two main disadvantages: (i) they support only single-stream transmissions (2) their training overhead scales linearly with the number of users. To support multi-stream transmission, \cite{mmWave_Estimation_2013} proposes an efficient mmWave channel estimation algorithm by leveraging the sparse nature of the channel and \textit{adaptive} compressed sensing tools. This solution, however, did not solve the overhead scaling issue.

In this paper, we propose a simple downlink system operation for multi-user mmWave systems based on compressed sensing channel estimation and conjugate analog beamforming. Contrary to prior work in \cite{Wang1, .13c, chen2011multi, Multilevel, Tsang,mmWave_Estimation_2013}, the training overhead of the proposed channel estimation solution does not scale with the number of users. Hence, it is of special interest for multi-user mmWave systems. Under certain assumptions, we characterize a lower bound on the achievable rate of the proposed system operation as a function of the compressed sensing measurements. Simulation results show that the proposed algorithm requires much less training overhead compared with traditional lower-frequency solutions, while employing mmWave-suitable hardware.

We use the following notation: $\bA$ is a matrix, $\ba$ is a vector, and $a$ is a scalar. $\|\bA \|_F$ is the Frobenius norm of $\bA$, whereas $\bA^\mathrm{T}$, $\bA^*$, $\bA^{-1}$, are its transpose, Hermitian, and inverse, respectively. $\bbE\left[\cdot\right]$ denotes expectation and $\mathbbm{1}_{\{.\}}$ is an indicator.
\section{System Model} \label{sec:Model}

\begin{figure} [t]
\centerline{
\includegraphics[width=1\columnwidth, height=125pt]{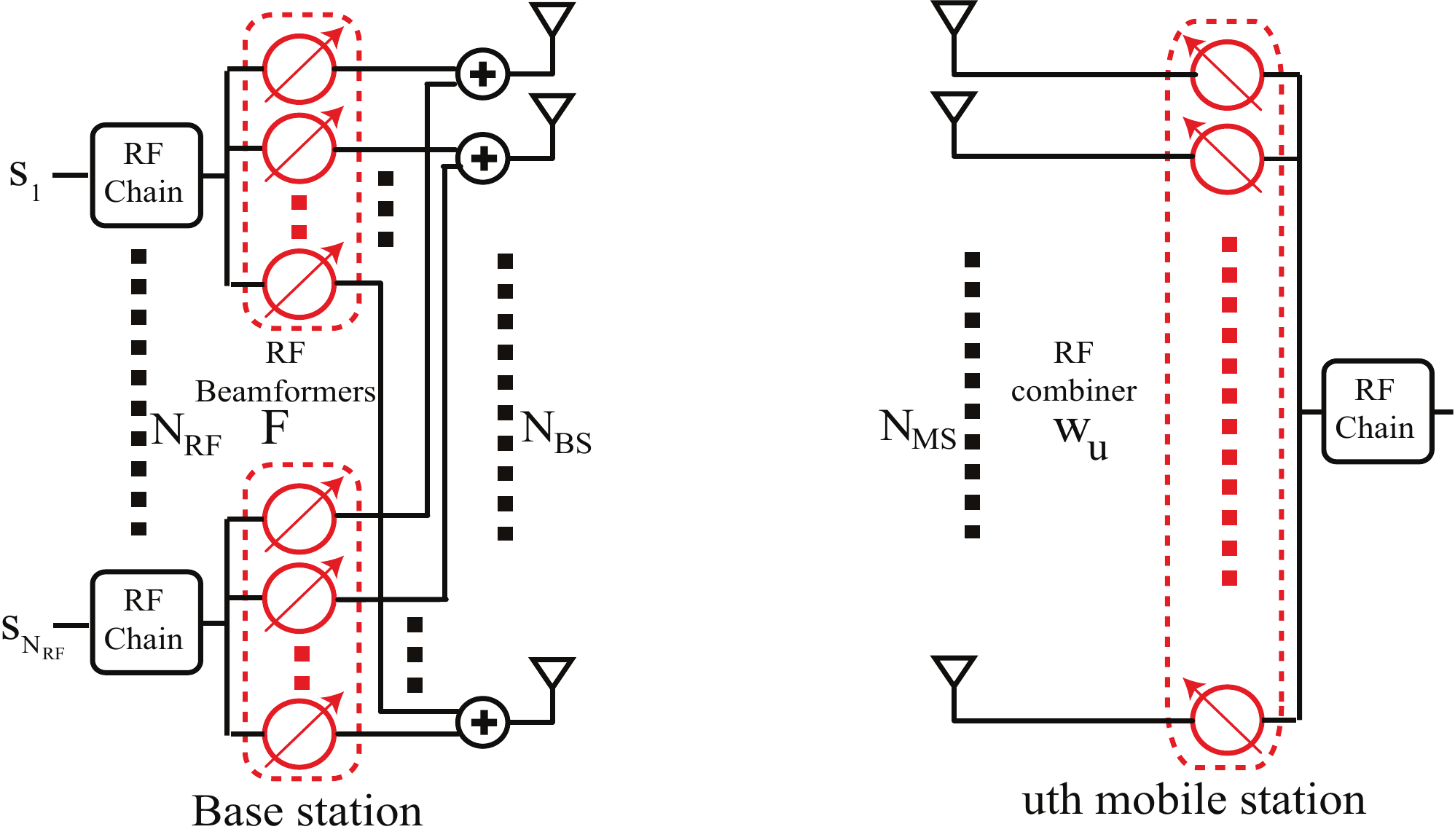}
}
\caption{A BS with RF beamformers and $N_\mathrm{RF}$ RF chains communicating with the $u$th MS that employs RF combining.}
\label{fig:Hybrid}
\end{figure}

Consider a mmWave system with a base station (BS) having $N_\mathrm{BS}$ antennas and $N_\mathrm{RF}$ RF chains as shown in  \figref{fig:Hybrid}. The BS is assumed to communicate with $U$ mobile stations (MS's), and each MS is equipped with $N_\mathrm{MS}$ antennas. We focus on the multi-user beamforming case in which the BS communicates with every MS via \textit{only one stream}. Further, we assume that the maximum number of users that can be simultaneously served by the BS equals the number of BS RF chains, i.e., $U \leq N_\mathrm{RF}$. This is motivated by the spatial multiplexing gain of the described multi-user precoding system, which is limited by $\min\left(N_\mathrm{RF},U\right)$ for $N_\mathrm{BS} > N_\mathrm{RF}$. For simplicity, we will also assume that the BS will use $U$ out of the $N_\mathrm{RF}$ available RF chains to serve the $U$ users.

On the downlink, the BS applies an $N_\mathrm{BS} \times U$ RF precoder, $\bF=\left[\bff_1, \bff_2, ..., \bff_U\right]$. The sampled transmitted signal is therefore $\bx= \bF \bs $, where $\bs=[s_1, s_2, ..., s_U]^\mathrm{T}$ is the $U \times 1$ vector of transmitted symbols, such that $\bbE\left[\bs\bs^*\right] = \frac{P_\mathrm{T}}{U}\bI_U$, and $P_\mathrm{T}$ is the average total transmitted power. Since $\bF$ is implemented using quantized analog phase shifters, $\left[\bF\right]_{m,n}= \frac{1}{\sqrt{N_\mathrm{BS}}} e^{\j \phi_{m,n}}$, where $\phi_{m.n}$ is a quantized angle, and the factor of  $\frac{1}{\sqrt{N_\mathrm{BS}}}$ is for power normalization.

For simplicity, we adopt a narrowband block-fading channel model \cite{ayach2013spatially,brady2013beamspace,mmWave_Estimation_2013}, by which the $u$th MS receives the signal
\begin{equation}
\br_{u}=\bH_{u}\sum_{r=1}^{U}{ \bff_{r} s_{r}} + \bn_{u},
\label{eq:received_signal}
\end{equation}
where $\bH_{u}$ is the $N_\mathrm{MS} \times N_\mathrm{BS}$ matrix that represents the mmWave channel between the BS and the $u$th MS, and $\bn_{u} \sim \cN (\boldsymbol{0}, \sigma^2 \bI )$ is a Gaussian noise vector.

At the $u$th MS, the RF combiner $\bw_u$ is used to process the received signal $\br_u$ to produce the scalar
\begin{equation}
y_u= \bw_u^*  \bH_{u}  \sum_{r=1}^{U}{\bff_{r} s_{r}} + \bw_u^* \bn_u,
\label{eq:combined_signal}
\end{equation}

MmWave channels are expected to have limited scattering~\cite{Rapp5G}. Therefore, and to simplify the analysis, we will assume a single-path geometric channel model \cite{Multilevel,barati2014directional}. Under this model, the channel $\bH_{u}$ can be expressed as
\begin{align}
\bH_u =\sqrt{N_\mathrm{BS} N_\mathrm{MS}} \alpha_{u} \ba_\mathrm{MS}\left(\theta_{u}\right) \ba^*_\mathrm{BS} \left(\phi_{u} \right),
\label{eq:channel_model}
\end{align}
where $\alpha_{u}$ is the complex path gain, including the path-loss, with $\mathbb{E}\left[|\alpha_{u}|^2\right]=\bar{\alpha}$. The variables $\theta_{u}$, and $\phi_{u} \in [0, 2\pi]$ are the angles of arrival and departure (AoA/AoD) respectively. Finally, $\ba_\mathrm{BS}\left(\phi_{u}\right)$ and $\ba_\mathrm{MS}\left(\theta_{u}\right)$ are the antenna array response vectors of the BS and $u$th MS respectively. The BS and each MS are assumed to know the geometry of their antenna arrays. While the results and insights developed in the paper can be generalized to arbitrary antenna arrays, we will assume uniform arrays in the simulations of Section \ref{sec:Results}.

\section{Proposed Downlink System Operation} \label{sec:Operation}
The proposed downlink operation for multi-user mmWave systems consists of two phases: (i) compressed sensing based downlink channel estimation and (ii) conjugate analog beamforming/combining. For the downlink channel training, random beamforming and projections are used to efficiently estimate the mmWave channel with relatively low training overhead thanks to the sparse nature of the channel. One main advantage of this technique is that all the MS's can simultaneously estimate their channels. Therefore, the training overhead does not scale with the number of users. This is contrary to the adaptive channel estimation and beamforming design techniques in \cite{mmWave_Estimation_2013,Multilevel,Wang1}, which are user-specific. The estimated channels are then used to build the analog beamformers and combiners. Extensions to hybrid analog/digital precoders are also possible \cite{alkhateeb2014limited}, but our focus in this paper is on the evaluation of compressed sensing channel estimation.

\subsection{Compressed Sensing Based Channel Estimation} \label{subsec:CS}
Given the geometric mmWave channel model in \eqref{eq:channel_model}, estimating the channel is equivalent to estimating the different parameters of the channel path; namely its AoA, AoD, and the complex gain. In this section, we exploit this poor scattering nature of the mmWave channel, and formulate the channel estimation problem as a sparse problem. We then briefly show how compressed sensing can be used to estimate the channel.

\textbf{A sparse formulation:} Consider the system and mmWave channel models described in \sref{sec:Model}. If the BS uses a training beamforming vector $\bp_m$, and the $u$th MS employs a training combining vector $\bq_n$ to combine the received signal, the resulting signal can be written as
\begin{equation}
y_{n,m}={\bq^H_n} \bH_u \bp_m s_m+ {\bq^H_n} \bn_{n,m},
\end{equation}
where $s_m$ is the training symbol on the beamforming vector $\bp_m$, and we use $s_m= \sqrt{P}$, with $P$ the average power used per transmission in the training phase. If the BS employs $M_\mathrm{BS}$ such beamforming vectors $\bp_m, m=1,..., M_\mathrm{BS}$, at $M_\mathrm{BS}$ successive time slots, and the MS uses $M_\mathrm{MS}$ measurement vectors $\bq_n, n=1,2,...,M_\mathrm{MS}$ at $M_\mathrm{MS}$ successive instants to detect the signal transmitted over \textit{each} of the beamforming vectors, the resulting received matrix will be \cite{mmWave_Estimation_2013}
\begin{equation}
\bY_\mathrm{MS}=\sqrt{P} \bQ^H \bH_u \bP+\bN,
\end{equation}
where $\bQ=\left[\bq_1, \bq_2, ..., \bq_{M_\mathrm{MS}}\right]$ is the  $N_\mathrm{MS}\times M_\mathrm{MS}$ measurement matrix, $\bP=\left[\bp_1,\bp_2,..., \bp_{M_\mathrm{BS}}\right]$ is the BS $N_\mathrm{BS}\times M_\mathrm{BS}$ beamforming matrix, and $\bN$ is an $M_\mathrm{MS} \times M_\mathrm{BS}$ noise matrix.

To exploit the sparse nature of the channel, we first vectorize the resultant matrix $\bY_\mathrm{MS}$ as in \cite{mmWave_Estimation_2013} to get
\begin{align}
\by_\mathrm{MS} = \sqrt{P} \left(\bP^T \otimes \bQ^H \right) \left( \ba_\mathrm{BS}^*\left(\phi_u\right) \otimes \ba_\mathrm{MS}\left(\theta_u\right) \right) \alpha_u+\bv, \label{eq:vec_signal}
\end{align}

To complete the problem formulation, we assume that the AoDs, and AoAs are taken from a grid of $G_\mathrm{BS}$ and $G_\mathrm{MS}$ points, respectively. By neglecting the grid quantization error, we can approximate $\by_\mathrm{MS}$ in \eqref{eq:vec_signal} as
\begin{align}
\by_\mathrm{MS} &= \sqrt{P} \left(\bP^T \otimes \bQ^H \right) \left(\overline{\bA}_\mathrm{BS}^* \otimes \overline{\bA}_\mathrm{MS}\right) \bz_u+\bv,\label{eq:sparse_formulation}
\end{align}
where the $N_\mathrm{BS} \times G_\mathrm{BS}$ matrix $\overline{\bA}_\mathrm{BS}$ and $N_\mathrm{MS} \times G_\mathrm{MS}$ matrix $\overline{\bA}_\mathrm{MS}$ are the dictionary matrices that consist of the column vectors $\ba_\mathrm{BS} \left(\bar{\phi}_k\right)$, and $\ba_\mathrm{MS}\left(\bar{\theta}_\ell\right)$, respectively, with $\bar{\phi}_k$, and $\bar{\theta}_\ell$ the $k$th, and $\ell$th points of the angle grids. $\bz_u$ is a $G_\mathrm{BS} G_\mathrm{MS} \times 1$ vector which carries the path gains of the corresponding quantized directions. Hence, $\bz_u$ is a sparse vector with only $1$ non-zero element. Note that detecting the column of $\overline{\bA}_\mathrm{BS}$ and $\overline{\bA}_\mathrm{MS}$ that corresponds to this non-zero element directly implies the detection of the AoA and AoD of the channel path.

\textbf{Compressed sensing measurements:} Thanks to the sparse formulation of the mmWave channel estimation problem in \eqref{eq:sparse_formulation}, compressed sensing tools can be leveraged to design efficient training beamforming/combining matrices \cite{mmWave_Estimation_2013,Ramasamy}. Considering the measurement matrix $\boldsymbol\Phi=\bP^T \otimes \bQ^H $, and the dictionary $\boldsymbol\Psi=\overline{\bA}_\mathrm{BS}^* \otimes \overline{\bA}_\mathrm{MS}$, one interesting research direction is to study the conditions on $\boldsymbol\Phi, \boldsymbol\Psi$ under which the support of the sparse vector $\bz_u$ can be recovered with high probability and with low training overhead. Leaving this objective for future work, we will try in this paper to get some insights into a sufficient (not necessarily the minimum) number of measurements, and the relation between this number and the achievable rate of mmWave systems. For that, we consider the following measurement matrix.

The BS will design its training beamforming matrix $\bP$, such that $\left[\bP\right]_{m,n}=e^{\j \phi_{m,n}}$ where $\phi_{m,n}$ is randomly and uniformly selected from the set of quantized angles $\{0,\frac{2 \pi}{N_Q^\mathrm{BS}}, ... , \frac{ \left(N_Q^\mathrm{BS}-1\right) 2\pi}{N_Q^\mathrm{BS}}\}$. Each MS similarly designs its training combining matrix $\bQ$, with $N_Q^\mathrm{MS}$ angle quantization bits. Note that for this design, each entry of the measurement matrix $\boldsymbol\Phi$ will also be equal to $e^{\j \zeta}$, with the angle $\zeta$  selected randomly from a certain quantized angle set.

\textbf{AoA/AoD estimation:} To estimate its channel AoA/AoD, each MS $u$ needs to recover the support of its sparse vector $\bz_u$. As the measurement and dictionary matrices $\boldsymbol\Phi, \boldsymbol\Psi$ are known for all MS's, each MS can use sparse recovery algorithms (such as LASSO \cite{LASSO}, Orthogonal Matching Pursuit (OMP) \cite{OMP}, etc.) to estimate its channel AoA/AoD. In the simulations of \sref{sec:Results}, we adopt OMP for low-complexity. In this case, the support of $\bz_u$, supp($\bz_u$), is determined by solving
\begin{equation}
\mathrm{supp}(\bz_u)=\arg\max{{\boldsymbol\Psi}^H {\boldsymbol\Phi}^H \by_\mathrm{MS}},
\end{equation}
which directly determines the estimated AoA/AoD ,$\hat{\theta_u}, \hat\phi_u$.
\subsection{Conjugate Analog Beamforming} \label{subsec:BF}
Given the estimated AoA, each MS $u$ will design its analog combining vector such that $\bw_u=\ba_\mathrm{MS}\left(\hat\theta_u\right)$. Each MS $u$ will also feed the index of its estimated AoD back to the BS which needs $\log_2 G_\mathrm{BS}$ bits. Finally, the BS designs its analog beamforming matrix $\bF$ to match the effective channels (including the effect of the combining vectors), i.e., the BS sets $\bF=\left[\ba_\mathrm{BS}\left(\hat\phi_1\right), \ba_\mathrm{BS}\left(\hat\phi_2\right), ..., \ba_\mathrm{BS}\left(\hat\phi_U\right)\right]$.
\section{Achievable and Effective Rates} \label{sec:Rates}
In this section, we evaluate the achievable rate of the proposed downlink mmWave system operation in a special case and try to get some insights into the relation between the performance of mmWave systems and compressed sensing measurements in more general cases.


Consider the system model in \sref{sec:Model} and the proposed conjugate analog beamforming/combining in \sref{subsec:BF}. For tractability, we make the following assumptions
\begin{assumption}
All path gains are constants. This is relevant to mmWave LOS paths, which are dominant in dense mmWave networks \cite{Bai1}.
\label{asu1}
\end{assumption}
\begin{assumption}
The BS and MS's employ uniform arrays.
\label{asu2}
\end{assumption}
\begin{assumption}
The sizes of the grids in \eqref{eq:sparse_formulation} are $G_\mathrm{BS}=N_\mathrm{BS}$, $G_\mathrm{MS}=N_\mathrm{MS}$, and the angle grid points $\bar\phi_k, \bar\theta_\ell,$ of the dictionary $\boldsymbol\Psi$ are the virtual directions satisfying $\frac{2 \pi d}{\lambda}\sin\left(\bar{\phi}_{k}\right)=\frac{2 \pi k}{N_\mathrm{BS}}$ and $\frac{2 \pi d}{\lambda}\sin\left(\bar{\theta}_{\ell}\right)=\frac{2 \pi \ell}{N_\mathrm{MS}}$\cite{Virtual}.
\label{asu3}
\end{assumption}

Under the assumptions \ref{asu2}-\ref{asu3}, the matrices $\overline{\bA}_\mathrm{BS}$ and $\overline{\bA}_\mathrm{MS}$ become DFT matrices \cite{Virtual}. Further, using the virtual channel model transformation in \cite{Virtual}, we note that the steering vectors of both the actual and estimated AoAs/AoDs (of the dictionary $\boldsymbol\Psi$) are columns of these matrices. Therefore, the rate of user $u$, $R_u$, can be written as
\begin{equation}
\log_2\left(1+\frac{ \mathbbm{1}_{\{\hat\theta_u=\theta_u, \phi_u=\hat\phi_u\}}} { {\mathbbm{1}_{\{\hat\theta_u=\theta_u\}}  \sum_{r=1}^{U} \mathbbm{1}_{\{ \phi_u=\hat\phi_r\}}+\frac{1}{\frac{\mathsf{SNR}}{U} N_\mathrm{BS} N_\mathrm{MS} \left|\alpha_u\right|^2}}}\right).
\end{equation}

Denoting $\mathring{R}_u=\log_2\left(1+\frac{\mathsf{SNR}}{U}N_\mathrm{BS} N_\mathrm{MS} \left|\alpha_u\right|^2\right)$ as the single-user rate (without interference), the average achievable rate of user $u$, $\bar{R}_u$, can be then lower bounded as
\begin{align}
\bar{R}_u & \stackrel{}{\geq} \bbE\left[\mathring{R}_u \mathbbm{1}_{\{ \bigcap_{r \neq u} \left(\phi_u \neq \hat\phi_r \right)\}} \mathbbm{1}_{\{\hat\theta_u=\theta_u, \phi_u=\hat\phi_u\}} \right], \\
& \stackrel{(a)}{=} \mathring{R}_u \mathrm{P}_{\{ \bigcap_{r \neq u} \left(\phi_u \neq \hat\phi_r \right)\})} \mathrm{P}_{\{\theta_u=\theta_u, \phi_u=\hat\phi_u\}},\\
& \stackrel{}{\geq} \mathring{R}_u  \left(1- \frac{U}{N_\mathrm{BS}}\right) \mathrm{P}_{\{\hat\theta_u=\theta_u, \phi_u=\hat\phi_u\}},
\end{align}
where (a) follows from the independence between the estimation success event $\{\hat\theta_u=\theta_u, \phi_u=\hat\phi_u\}$ and the single-user rate given assumption \ref{asu1}. Now, we note that $\mathrm{P}_{\{\hat\theta_u=\theta_u, \phi_u=\hat\phi_u\}}$ is the probability of correct support recovery of the sparse vector $\bz_u$. This directly relates the achievable rate to the compressed sensing literature. For a general relation, assume the channel fading coherence equals $L_C$ symbols, the effective achievable rate of user $u$ (considering the training overhead) can be written as
\begin{equation}
\bar{R}_{u,\mathrm{eff}} \geq \mathring{R}_u  \left(1- \frac{U}{N_\mathrm{BS}}\right) \left(1-\frac{M_\epsilon}{L_C}\right) (1-\epsilon),
\label{eq:Imp}
\end{equation}
where $M_\epsilon$ equals the number of measurements needed to guarantee the  support recovery of the sparse vector $\bz_u$ with probability at least $1-\epsilon$. Given this relation in \eqref{eq:Imp}, it is interesting to design the measurement matrix $\boldsymbol\Phi$ to maximize this lower bound on the achievable rate. It is also of interest to optimize the number of measurements to handle the tradeoff shown between the accurate channel estimate (small $\epsilon$) and large training overhead.
 \begin{figure} [t]
\centerline{
\includegraphics[width=1.1\columnwidth]{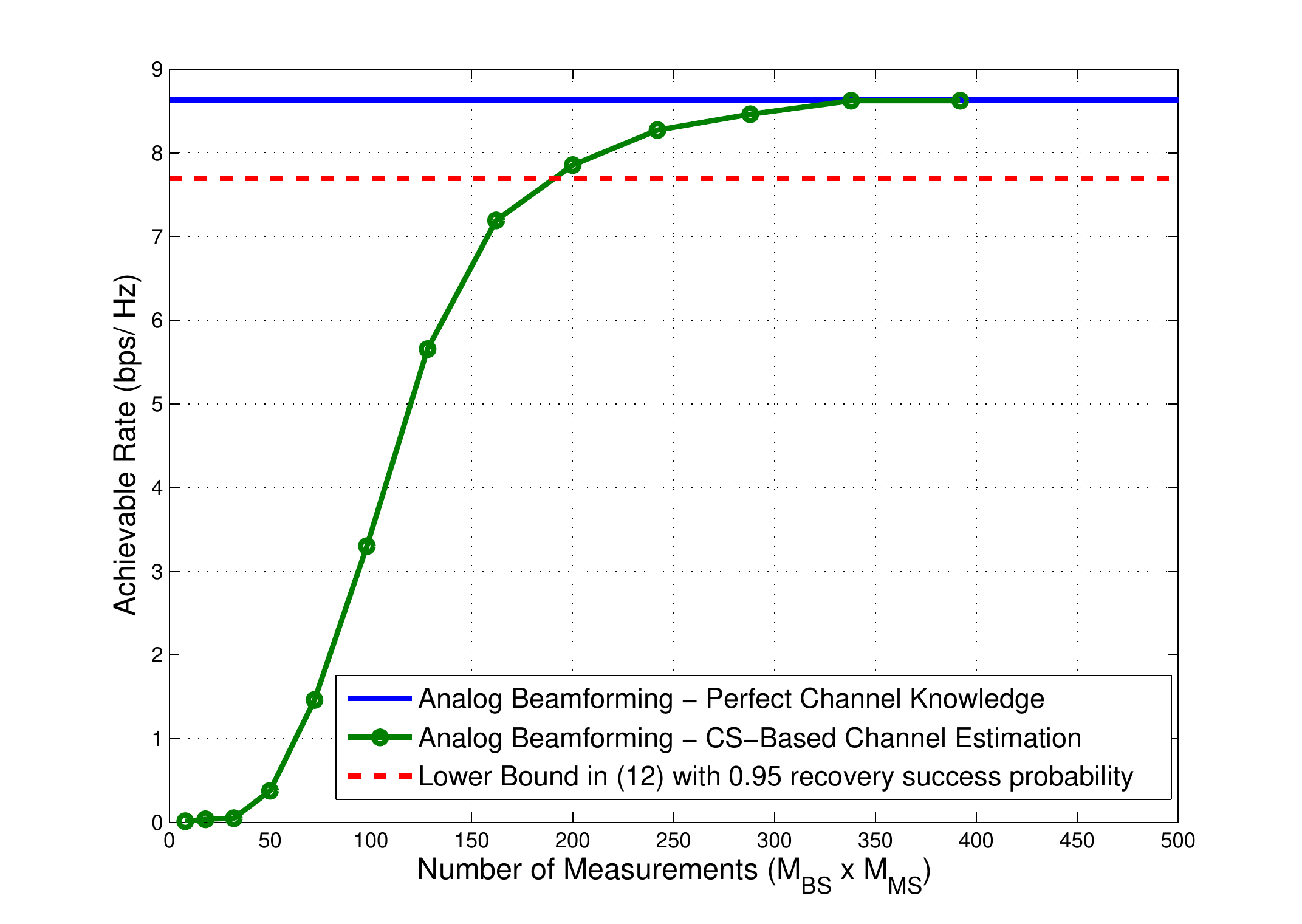}
}
\caption{Achievable rates using the proposed system operation for different numbers of compressed sensing measurements }
\label{fig:Num}
\end{figure}
\section{Simulation Results} \label{sec:Results}
In this section, we evaluate the performance of the proposed downlink mmWave system operation. We consider the system model in \sref{sec:Model} with the BS employing a ULA of $64$ antennas and $4$ MS's, each with $32$ antennas. The system operates at $28$ GHz with a bandwidth of $50$ MHz. The BS-MS distance is $500$ m, and all channels are LOS and single-path. In the channel estimation phase, the BS and MS's apply the random beamforming and measurement procedure described in \sref{subsec:CS} with $N_Q^\mathrm{BS}=N_Q^\mathrm{MS}=4$, and with average transmit power equal to $37$ dBm. The beamformers/combiners are then built as described in \sref{subsec:BF}.

In \figref{fig:Num}, the per-user achievable rate of the proposed system operation is shown versus the number of compressed sensing measurements. This figure illustrates that $\sim 280-330$ measurements are needed to approach the achievable rate with perfect channel knowledge. While this number may look large, it is actually an order of magnitude less than what is required  by traditional lower frequency solutions, which is $ \sim N_\mathrm{BS} N_\mathrm{MS} = 2048$ symbols. Note also that this training overhead does not depend on the number of users, which makes compressed sensing of special interest to multi-user mmWave systems. Compared with adaptive compressed sensing techniques \cite{mmWave_Estimation_2013}, they may require less training to estimate each user channel, e.g., $\sim 50-100$ in \cite{mmWave_Estimation_2013}. This overhead, however, scales with the number of users, which means $~ 200-400$ for $4$ users, and larger if more users are served. Finally, note that these results just illustrate a sufficient number of measurements given the proposed design of the measurement matrix $\boldsymbol\Phi$ in \sref{subsec:CS}. Therefore, optimizing this design may lead to even smaller training overhead.

\begin{figure} [t]
\centerline{
\includegraphics[width=1.1\columnwidth]{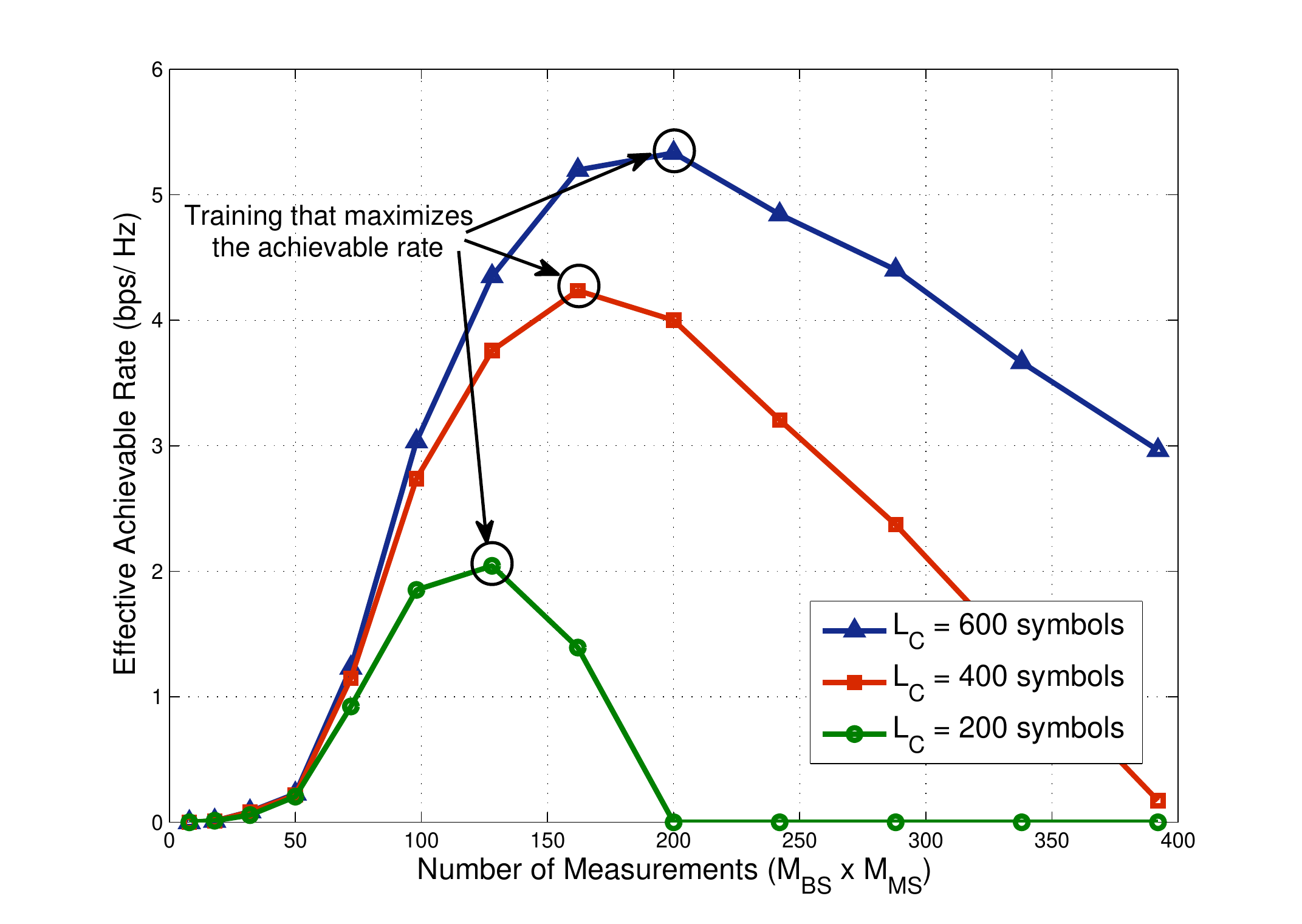}
}
\caption{Achievable rates using the proposed system operation for different values of channel fading coherence.}
\label{fig:Opt}
\end{figure}

In \figref{fig:Opt}, the effective achievable rate is shown for different numbers of fading coherence values, $L_C$. This figure indicates that it is important to compromise between the accuracy of the channel estimate and the training overhead to maximize the achievable rate, especially with fast channels.

\section{Conclusion}
In this paper, we proposed and evaluated a low-complexity downlink system operation for multi-user mmWave systems based on compressed sensing channel estimation. Simulation results showed that the proposed system operation requires a relatively small training overhead (w.r.t. the channel matrix dimensions) to achieve a very good performance. Results also indicated that the number of measurements need to be wisely selected to maximize the effective system sum-rate.

\clearpage
\begin{small}

\end{small}

\end{document}

%% file: input.tex
\usepackage{amsfonts}
\usepackage{times}
\usepackage{graphicx}
\usepackage{latexsym}
\usepackage{dsfont}
\usepackage{amssymb}
\usepackage{amsmath}
\usepackage{cite}
\usepackage{verbatim}
\usepackage{subfigure}
\newtheorem{theorem}{Theorem}

\newtheorem{assumption}[theorem]{Assumption}

\newcommand{\figref}[1]{{Fig.}~\ref{#1}}

\def\bb0{{\mathbb{0}}}

\def\ba{{\mathbf{a}}}
\def\bb{{\mathbf{b}}}

\def\bff{{\mathbf{f}}}

\def\bn{{\mathbf{n}}}

\def\bp{{\mathbf{p}}}
\def\bq{{\mathbf{q}}}
\def\br{{\mathbf{r}}}
\def\bs{{\mathbf{s}}}

\def\bv{{\mathbf{v}}}
\def\bw{{\mathbf{w}}}
\def\bx{{\mathbf{x}}}
\def\by{{\mathbf{y}}}
\def\bz{{\mathbf{z}}}
\def\b0{{\mathbf{0}}}

\def\bA{{\mathbf{A}}}

\def\bF{{\mathbf{F}}}

\def\bH{{\mathbf{H}}}
\def\bI{{\mathbf{I}}}

\def\bN{{\mathbf{N}}}

\def\bP{{\mathbf{P}}}
\def\bQ{{\mathbf{Q}}}

\def\bY{{\mathbf{Y}}}

\def\bbE{{\mathbb{E}}}

\def\cN{\mathcal{N}}

\def\sf0{{\mathsf{0}}}

